\documentclass[aps,prl,reprint,groupedaddress,showpacs]{revtex4-1}
\usepackage{graphicx}
\usepackage{dcolumn}
\usepackage{bm}
\usepackage{amssymb}
\usepackage{mathrsfs}
\usepackage{CJK}

\begin{document}

\title{A model independent study of nonlocality with polarization entangled photons}

\author{Boya Xie}
\author{Sheng Feng}
\email[]{fengsf2a@hust.edu.cn}

\affiliation{Hubei Key Laboratory of Modern Manufacturing Quality Engineering, School of Mechanical Engineering, Hubei University of Technology, Wuhan 430068, China}

\date{\today}



\begin{abstract}
Nonlocality as a fundamental aspect of quantum mechanics is witnessed by violation of Bell inequality or its variants, for which all relevant studies assume some correlations exhibited by local realistic theories. The strategy of Bell's theorem is to establish some criteria to distinguish local realistic theories from quantum mechanics with respect to the nonlocal nature of entangled systems. Here we propose a model independent study of nonlocality that needs not to assume any local theory since observation of the expected nonlocal effect is straightforward quantum mechanically. Our proposal involves a bipartite polarization-entangled system in which one photon immediately reduces into a circular-polarization (CP) state when its partner at a space-like distance is detected in another CP state. The state reduction of the photon can be mechanically monitored because a CP photon carries angular momentum and exerts a torque on a half-wave plate whose mechanical motion is measurable, which is well described by quantum mechanics and independent of any local realistic assumption.
\end{abstract}

\maketitle

\noindent Eight decades ago, one of the most far-reaching debates in scientific history occurred between Bohr and Einstein on the completeness of quantum mechanical description of physical reality \cite{Einstein1935,Bohr1935}. It was the statistical nature of quantum mechanics (QM) that first caught the attention of Einstein who believed that quantum systems were controlled by hidden variables that determined measurement outcomes. Later in 1935, Einstein, Podolsky, and Rosen (EPR) brought the nonlocal feature of QM into the public sight \cite{Einstein1935}, stimulating many attempts \cite{Bohm1952,Bohm1957,Bell1964,Clauser1969,Pitowsky1982,Greenberger1990,Hardy1993,Wiseman2007} to explore the EPR paradox for a satisfactory solution.

A turning point of the long-lasting debate came in 1964 when Bell published an inequality \cite{Bell1964} giving an upper bound on the strength of some correlations exhibited by local realist theories. According to Bell's theorem, violation of the inequality would conclusively preclude all local realistic theories. But early Bell experiments \cite{Freedman1972,Aspect1982} were performed under imperfect conditions and forced to make additional assumptions to deny local realism \cite{Genovese2005}. Long after those pioneering works, three ``loophole-free" Bell experiments were reported at last in 2015 \cite{Giustina2015,Shalm2015,Hensen2015}.

Bell's theorem, together with Bell experiments, has deeply influenced our perception and understanding of physics and is essential for the applications of quantum information technologies. Quantum nonlocal correlations, as witnessed by the violation of Bell inequality, are now thought as a fundamental aspect of quantum theory by most physicists \cite{Walborn2011,Cavalcanti2011,Christensen2013,Hirsch2013,Erven2014,Brunner2014,Popescu2014,Aspect2015}. Yet complete consensus has not been reached in the literature over whether the door has been closed on the Bohr-Einstein debate \cite{Aspect2015,Khrennikov2015,Kupczynski2017,Pons2017}. In the spectrum of opinions on the results of Bell experiments, still on the focus of the question is quantum nonlocality whereby particles appear to influence one another instantaneously \cite{Orlov2002,Matzkin2008,Khrennikov2015,Kupczynski2017,Pons2017}.

\begin{figure}[htbp]
\centering
\includegraphics[width=7cm]{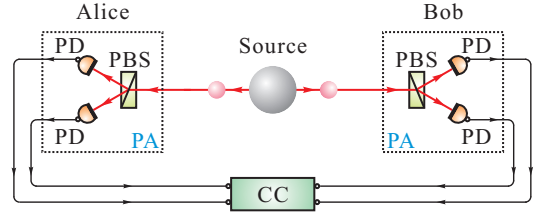}
\caption{(color online)  A typical Bell experiment on the nonlocal nature of polarization entangled photons, which is evidenced by correlations between the local measurement results of Alice and Bob who are widely separated in space. PD: Photon detector. PBS: Polarizing beamsplitter. PA: Polarization analyzer. CC: Coincidence counter.}
\label{fig:bell}
\end{figure}

In a typical configuration of Bell experiments, quantum nonlocality is usually demonstrated with polarization entangled photons as shown in Fig. \ref{fig:bell}. Suppose that the two photons shared by Alice and Bob are in an entangled state described by
\begin{eqnarray}\label{eq:ent1}
|\psi\rangle&=&2^{-1/2}(|H\rangle_1|H\rangle_2+|V\rangle_1|V\rangle_2),
\end{eqnarray}
a superposition of a $|H\rangle_1|H\rangle_2$ (both photons horizontally polarized) state and a $|V\rangle_1|V\rangle_2$ (both vertically polarized) state. According to QM, the polarization states of both photons are undetermined unless one of them is measured by a polarization detector and irreversibly projected into an eigenstate of an operator describing the measurement. For example, if Alice's photon is in a $|H\rangle$ (or $|V\rangle$) state after the measurement, it follows from Eq. (\ref{eq:ent1}) that then Bob's photon will immediately collapse into a $|H\rangle$ (or $|V\rangle$) state too, or vice versa, due to the nonlocal correlation between the two photons no matter how widely they are separated in space. This nonlocality owned by the entangled photons is referred to as ``spooky action at a distance" by Einstein.

It is Bell's theorem, as a widely-accepted theoretical model, that for the first time paved a way to distinguish experimentally from local realistic theories quantum mechanics that predicts nonlocality for complex systems described by Eq. (\ref{eq:ent1}). All experiments based on Bell's theorem (Bell experiments) and their variants share two common features with later investigations on quantum nonlocality: (1) Model dependence and (2) Lorentz invariance \cite{Hardy1992}. The model dependence comes from the fact that all relevant studies must first assume some local realist model that is then rejected according to some criteria \cite{Bell1964,Greenberger1990,Hardy1993,Wiseman2007}, whereas the feature of Lorentz invariance is partially related to the lack of spatiotemporal information in the wave function of the complex system. In this paper, we will show that it is possible to directly observe the nonlocal property of two entangled photons, which assumes no local realist theory and hence is model independent. The model independence guarantees that the experiment, if successfully conducted, will be a more convincing test of quantum nonlocality.

The key idea is that entanglement-induced state collapse of a photon may cause directly observable effects, such as angular momentum transfer between light and matter, at a space-like distance. To appreciate this insight, let rewrite Eq. (\ref{eq:ent1}) as
\begin{eqnarray}\label{eq:angular}
|\psi\rangle&=&2^{-1/2}(|L\rangle_1|R\rangle_2+|R\rangle_1|L\rangle_2),
\end{eqnarray}
in which $|L\rangle\equiv(|H\rangle-i|V\rangle)/\sqrt{2}$ and $|R\rangle\equiv(|H\rangle+i|V\rangle)/\sqrt{2}$ are the left-hand and right-hand CP states, respectively. It follows from Eq. (\ref{eq:angular}) that, when Alice's photon is detected in a $|L\rangle$ (or $|R\rangle$) state, Bob's photon even at a space-like distance will collapse immediately into a $|R\rangle$ (or $|L\rangle$) state \cite{Pan2000} that carries angular momentum \cite{Beth1936}, or vice versa.

\begin{figure}[htbp]
\centering
\includegraphics[width=7cm]{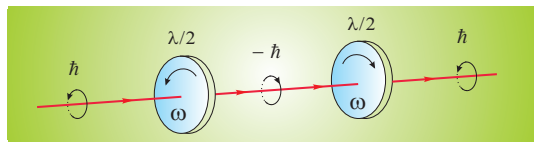}
\caption{(color online)  A mechanical detector for detection of the CP state of a photon, which changes the rotational speed of each half-wave plate by $\Delta\omega=\pm 2\hbar/I$ ($I$ is the moment of inertia) \cite{Beth1936}, with the sign of $\Delta\omega$ depending on the CP state of the incident photon. In contrast, a photon in a linear-polarization (LP), i.e., $|H\rangle$ or $|V\rangle$, state carries no angular momentum and cannot change the mechanical motion states of the plates. One should stress that a blue shifted or red shifted CP photon at the output remains in the same CP state as the input after exiting from the detector.}
\label{fig:polana}
\end{figure}

Of essence in direct observation of quantum nonlocality is that, after state reduction, Bob's photon carrying angular momentum can exert a mechanical torque on a half-wave plate \cite{Beth1936}, which distinguishes this work from all previous investigations that must assume local realistic theories. Based on the physical process of angular momentum exchange between half-wave plates and CP photons, one may construct a mechanical detector consisting of two freely-rotating half-wave plates in a row (Fig. \ref{fig:polana}) to detect, but not to alter, the state of a CP photon: After a CP photon exerts a torque on a half-wave plate when passing through it, the angular momentum of the plate will change by twice that of the incident photon, i.e., $|l_i|=2\hbar$ ($i=1,2$ are the sequential numbers for the plates and $\hbar\equiv h/2\pi$, where $h$ is Planck constant), but the signs of $l_i$ are opposite. At the output of the second plate, the CP state of the photon remains the same as the initial at the input of the first plate, but the state information of the photon is recorded by the mechanical detector that outputs a corresponding value of $\Delta l=\pm4\hbar$ ($\Delta l \equiv l_2-l_1$). The sign of $\Delta l$ reveals the specific CP state of Bob's photon, into which the photon collapses as a result of detecting Alice's photon in another CP state at a space-like distance.

Yet there is a subtlety here that needs to be clarified as follows: In its interaction with the CP photons, each wave plate gains angular momentum that must be accompanied by nonzero energy exchange between the photon and the plate. Although the final CP state of the photon is the same as the initial, it gains or loses part of its energy when it exerts a torque on each plate. Consequently, the wavelength of the photon is blue shifted or red shifted after it exits from the second plate, with its CP state unchanged though.

It then follows that the core elements in the direct observation of the nonlocal nature of entangled photons can be fully described in the theoretical framework of QM: (a) Bob's photon at a space-like distance is immediately projected into a CP state by the detection of Alice's photon in another CP state according to Eq. (\ref{eq:angular}) \cite{Pan2000}; (b) A photon in a CP state changes the angular motion of a half-wave plate by exerting a mechanical torque on it \cite{Beth1936}. Therefore, the expected results if achieved successfully in experiment will be a model independent evidence for quantum nonlocality since no local realist model is needed.


However, the relevant physical effect may be too small to observe, i.e., it is technically intractable to monitor the angular momentum change of a wave plate caused by a single photon. In what follows, we will discuss how to implement a direct observation of nonlocality in experiment with currently available technologies. Firstly, one should note that many photons per second can be generated with bright photon-pair sources, facilitating the observation of the expected physical effect. Secondly, in light of technical challenges in experiment, one may utilize an optical amplifier to boost the power of a photon-bearing beam before it enters into the mechanical detector.

\begin{figure}[htbp]
\centering
\includegraphics[width=8cm]{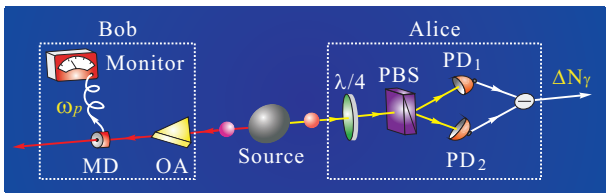}
\caption{(color online) Schematics for direct observation of the nonlocal correlation between entangled photons, through mechanical detection of the angular momentum exchange between CP photons and matter. MD: Mechanical detector. OA: Optical amplifier. $\lambda/4$: 1/4-wave plate. PBS: Polarizing beamsplitter. PD$_{1,2}$: Photon detectors.}
\label{fig:exp1}
\end{figure}

In the following, we will give a quantitative estimation of the signal size in direct observation of nonlocality with polarization entangled photons. To this aim, let suppose that one has a light source, emitting $N_\gamma$ photon pairs per second in the state described by Eq.(\ref{eq:angular}), and an optical amplifier with a power gain of $G$ (Fig. \ref{fig:exp1}). One photon (Alice's photon) of each pair is projected into a $|L\rangle$ or $|R\rangle$ state by a CP analyzer that consists of a 1/4-wave plate, a polarizer, and a photon detector \cite{Pan2000}. Then the other twin (Bob's photon) at a space-like distance will immediately undergo a state reduction into a $|R\rangle$ or $|L\rangle$ state and be directed to a power amplifier at whose output one has $G$ photons in the same state as that at the amplifier's input. Therefore, the total number of photons entering into a mechanical detector during a time of $t$ second is $GtN_\gamma$, with roughly $GtN_\gamma/2$ photons in each CP state.

Although the angular momentum change ($2\hbar$) of each plate in the mechanical detector caused by an incident $|L\rangle$ photon cancels out that (-2$\hbar$) by a $|R\rangle$ photon, the number of $|L\rangle$ photons is usually unequal to that of $|R\rangle$ photons during any finite period of time $t$, thanks to quantum fluctuations. The photon number difference between $|L\rangle$ and $|R\rangle$ states is of the order of $\sqrt{tN_\gamma}$ before power amplification, after which time the number difference becomes $G\sqrt{tN_\gamma}$. Hence, the relative angular momentum change of the two wave plates can be calculated as of the order of
\begin{eqnarray}\label{eq:DJ}
\Delta J = 4\hbar \cdot G \sqrt{tN_\gamma}.
\end{eqnarray}
One should note that the above estimations for photon number fluctuations suffice for the purpose of this work, albeit they are not accurate since the photons are not in coherent states where minimal photon number fluctuations occur.

\begin{figure}[htbp]
\centering
\includegraphics[width=6cm]{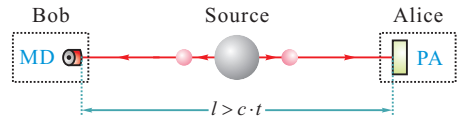}
\caption{(color online) Nonlocality condition $l>c\cdot t$ that must be satisfied by the measurements performed by Alice and Bob who are separated by a distance of $l$. This condition guarantees that any disturbance caused by Alice's measurement for $\Delta N_\gamma$ cannot propagate to Bob until his measurement for $\Delta\omega_p$ is finished, and vice versa. $c$ is the speed of light in vacuum and $t$ is the time interval during which both Alice and Bob carry out their measurements. MD: Mechanical detector. PA: Polarization analyzer.}
\label{fig:nlccon}
\end{figure}

For a disk-like wave plate, the moment of inertia is
\begin{eqnarray}\label{eq:moi}
I_m&=&(\pi/2)\rho D r^4,
\end{eqnarray}
wherein $\rho$ is the mass density, and $D$ stands for the plate thickness with $r$ being the radius of the disk bottom. Then it follows from Eqs. (\ref{eq:DJ}) and (\ref{eq:moi}) that the variation of the relative angular velocity of the two half-wave plates in the mechanical detector caused by the total torque of $GtN_\gamma$ CP photons is
\begin{eqnarray}\label{eq:as}
\Delta \omega_p&=&\frac{\Delta J}{I_m}=\frac{4\hbar G\sqrt{tN_\gamma}}{I_m}=\frac{8\hbar G}{\pi\rho D r^4}\Delta N_\gamma,
\end{eqnarray}
where $\Delta N_\gamma\equiv \sqrt{tN_\gamma}$. Eq. (\ref{eq:as}) shows a linear relationship between $\Delta\omega_p$ and $\Delta N_\gamma$, which means that the output of the mechanical detector is proportional to (correlated with) the photon number difference between the two CP states detected by Alice at a space-like distance. Therefore, Eq. (\ref{eq:as}) signifies nonlocality provided that the measurement speed $\propto t^{-1}$ is high enough to satisfy nonlocality condition \cite{Genovese2005} (Fig. \ref{fig:nlccon}).

\begin{figure}[htbp]
\centering
\includegraphics[width=3.0cm]{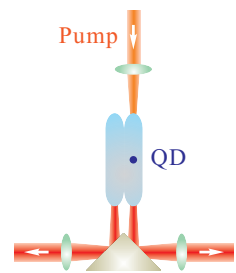}
\caption{(color online) A bright light source for polarization-entangled photon-pair generation. The photon pairs are produced by a quantum dot, which is coupled to one of two identical microcavities \cite{Dousse2010}. QD: Quantum dot.}
\label{fig:source}
\end{figure}

From Eq. (\ref{eq:as}), one may estimate an achievable magnitude of $\Delta\omega_p$ with existing technologies. Let use an ultrabright source of entangled photon pairs \cite{Dousse2010}, emitting $N_\gamma = 10^{9}$ photon pairs per second (Fig. \ref{fig:source}), and the photon pairs are entangled in polarization. One photon-bearing beam from the source is fed into an optical amplifier with cavity-enhanced cascaded stages, providing a total gain of $G=10^6$. If the beam wavelength is $\lambda=0.8\ \mu$m, then the incident light into the mechanical detector is of the order of $G N_\gamma E_\gamma = 2.5$ mW ($E_\gamma$ is the photon energy).

As for the sizes of the wave plates, the lower bound of their thickness is set by the birefringence ($\Delta n$) of the material. If fabricated with some birefringent material akin to Potassium Titanyl Phosphate (KTiOPO$_4$, or KTP), a zero-order half-wave plate of this kind will have a thickness $D = \lambda/(2\Delta n) \approx 4.5 \ \mu$m and a mass density of $3\times 10^3$ kg/m$^3$. For a designed bottom radius of $r = 50 \ \mu$m, one may obtain from Eq. (\ref{eq:as}) roughly $\Delta\omega_p \approx 1.8\times10^{-6}$ radius/s within a period of $t = $ 0.3 ms, during which time a signal at the speed of light travels through a distance of about 90 km in the air.

From the above estimations, one may design an experiment on direct observation of the nonlocal correlation between entangled photons as follows (Fig. \ref{fig:exp1}): Twin beams carrying entangled photons are produced by a cavity-coupled quantum dot. One beam is sent to Alice's CP analyzer \cite{Pan2000} with its twin sent into a mechanical detector after being amplified. Alice's analyzer and Bob's mechanical detector are separated by more than 90 km (Fig. \ref{fig:nlccon}) and, if the experiment is ground based, the twin beams may transport via low-noise fibers to arrive their destinations.

During the experiment, the light source is turned on for a period of $t = $ 0.3 ms such that Alice and Bob perform their local measurements for $\Delta N_\gamma$ and $\Delta\omega_p$ respectively. The photons sent to Alice are projected into $|L\rangle$ or $|R\rangle$ states and, immediately, the photons flying towards Bob collapse into corresponding CP states according to Eq. (\ref{eq:angular}). Bob's photons are first directed into the power amplifier and then the amplified light beam carrying CP photons enters into the mechanical detector where it transmits through the wave plates, whose relative angular speed consequently changes by a magnitude of $\Delta\omega_p$. In case that the photon flux exceeds the detection ability of photon counters in use, and one may instead utilize quantum-noise-limited heterodyne detector \cite{xie2018} to measure the light-field amplitude to obtain $\Delta N_\gamma$. After the 0.3 ms period of time, the light source is turned off so that the wave plates may freely rotate for say $\tau =$ 100 s during which time the relative angle of the two plates will change by of the order of $\theta=\Delta\omega_p \tau\approx 0.01^\circ$. The magnitude of $\theta$ is within the detection capability of a null polarimeter near shot noise limit \cite{he2015}.

The above experimental procedure can be repeated as long as allowed by experimental conditions and one may extract the temporal correlation between $\Delta\omega_p$ and $\Delta N_\gamma$ from data with
\begin{eqnarray}\label{eq:corram}
C_p&=&\frac{\sum_i \Delta\omega_p(i) \cdot \Delta N_\gamma(i)}{\sqrt{\sum_i |\Delta\omega_p(i)|^2} \cdot \sqrt{\sum_i |\Delta N_\gamma(i)|^2}},
\end{eqnarray}
wherein $i$ represents the sequential number of repeated data acquisition in the experiment. Ideally, $\Delta\omega_p$ and $\Delta N_\gamma$ should keep their relationship of Eq. (\ref{eq:as}), plugging which into Eq. (\ref{eq:corram}) leads to $C_p = 1$. Therefore, an experimental value of $C_p$ approaching unity is a signature for nonlocality. Otherwise, one should have $C_p = 0$.

Given that there might be a nonzero time delay between the data sets for $\Delta\omega_p$ and $\Delta N_\gamma$ in practice, one may define a generalized correlation function as,
\begin{eqnarray}\label{eq:corramg}
C_{\pm}(j)&=&\frac{\sum_i \Delta\omega_p(i) \cdot \Delta N_\gamma(i\pm j)}{\sqrt{\sum_i |\Delta\omega_p(i)|^2} \cdot \sqrt{\sum_i |\Delta N_\gamma(i)|^2}},
\end{eqnarray}
in which $j$ has the same meaning as that of $i$. Naively, one may expect $C_{\pm}(0)=C_p$, $C_{-}(j)=C_+(-j)$, and $C_{\pm}(j)|_{j\rightarrow \infty}=0$. Therefore, the amplitude of the correlation function $C_{\pm}(j)$ has a peak at $j=0$ and gradually decreases with $j$ until it reaches 0. For experiments with imperfect parameters, a peak at $j=0$ for $C_{\pm}(j)$ signifies nonlocality for the entangled photons.


In addition to the correlation functions $C_p$ and $C_{\pm}(j)$, there are other alternative ways to present experimental results for nonlocality. For instance, one may define
\begin{eqnarray}\label{eq:diff}
\Delta_{\pm}(i)=\frac{\Delta\omega_p(i)}{\sqrt{\sum_i |\Delta\omega_p(i)|^2}}
\pm\frac{\Delta N_\gamma(i)}{\sqrt{\sum_i |\Delta N_\gamma(i)|^2}},
\end{eqnarray} 
which may be thought of as a series of sequential sampling data points of some pseudo-time-varying signal $\Gamma_{\pm}(t)$,
\begin{eqnarray}\label{eq:pss}
\Delta_{\pm}(i) \equiv \Gamma_{\pm}(t)|_{t=i\cdot\iota}.
\end{eqnarray} 
Here $\iota$ stands for the real time interval between $\Delta\omega_p(i)$ and $\Delta\omega_p(i+1)$. From Eq. (\ref{eq:pss}) one may find squeezing in the noise of $\Gamma_{-}(t)$,
\begin{eqnarray}\label{eq:sqz}
\chi_s(\omega)={\int}^{+\infty}_{-\infty}{\rm d}\tau {\rm e}^{{\rm i}\omega \tau}<\Gamma _{-}(t) \Gamma_{-}(t+\tau)>.
\end{eqnarray}
When nonlocal correlation exists between $\Delta\omega_p(t)$ and $\Delta N_\gamma(t)$, one expects that $\chi_s(\omega)<\chi_a(\omega)$ where $\chi_a(\omega)$ stands for the noise of $\Delta_{+}(t)$,
\begin{eqnarray}\label{eq:sum}
\chi_a(\omega)={\int}^{+\infty}_{-\infty}{\rm d}\tau {\rm e}^{{\rm i}\omega \tau}<\Gamma _{+}(t) \Gamma_{+}(t+\tau)>.
\end{eqnarray}
In contrast, if there is no nonlocal correlation between $\Delta\omega_p(t)$ and $\Delta N_\gamma(t)$, one should have $\chi_s(\omega)=\chi_a(\omega)$.

To conclude, we have proposed a model independent study of quantum nonlocality with polarization entangled photons. The key idea is the mechanical detection of the CP state of a photon that carries angular momentum and exerts torques on half-wave plates. The plates' mechanical motion states vary depending on the specific CP state of the incident photon. We have provided a quantitative estimation of the expected signal size with existing technologies and suggested some measurable quantities as signatures for nonlocality owned by the twin photons such that experimental results may be unambiguously interpreted. Because no local realist model is assumed in the investigation, if the proposal is implemented experimentally, the results should be a model independent evidence for quantum nonlocality.

This work was supported by the National Natural Science Foundation of China (grant No. 11947134 and grant No. 12074110).




%

\end{document}